\documentstyle[11pt,epsfig]{article}
\begin{document}

\title{Spin Hall Effect}

\author{John Schliemann\\
Institute for Theoretical Physics, University of Regensburg,\\
D-93040 Regensburg, Germany}

\maketitle

\begin{abstract}
The intrinsic spin Hall effect in semiconductors has developed to
a remarkably lively and rapidly growing branch of research
in the field of semiconductor spintronics.
In this article we give a pedagogical overview on both theoretical and 
experimental accomplishments and challenges. Emphasis is put on
the the description of the intrinsic mechanisms of spin Hall transport
in III-V zinc-blende semiconductors,
and on the effects of dissipation.
\end{abstract}

\section{Introduction}

Starting from about the late 1990s, a pronounced, and partially also renewed,
interest in effects of spin-orbit coupling in semiconductors has been emerging
in the solid-state community. This development is mainly fueled by the
field of spintronics. The latter keyword summarizes an entire plethora
of theoretical and experimental efforts towards using the spin degree of 
freedom of electrons, instead, or in combination with, their charge for 
information processing, or, even more ambitious, for quantum information
processes. Thus, controlling the electron spin in semiconductor structures
is a key challenge, and the relativistic effect of spin-orbit coupling
is an important, if not indispensable, ingredient for reaching this goal.
Among all the rapidly progressing activities in this field,
a major and very recent development was the theoretical prediction 
and subsequent experimental investigation of the spin Hall effect in 
semiconductor structures. This effect amounts in a spin current, as opposed 
to a charge current, driven by a perpendicular electric field. In this  
article we review recent studies on spin Hall transport in semiconductors 
induced by intrinsic mechanisms.

A brief overview on important aspects and perspectives 
of the field of spintronics is given in the article by Wolf {\it et al.}
\cite{Wolf01}. Selected topics are reviewed in more detail in a volume
edited by Awschalom, Loss, and Samarth \cite{Awschalom02}. A comprehensive
review of many parts of spintronics was given by Zutic, Fabian, and Das
Sarma \cite{Zutic04}. Research directions not covered here include the
field of ferromagnetic semiconductors; for a review on this we refer to
Refs. \cite{Dietl02,Konig03,Dietl03,Timm03,Dietl04,MacDonald05}.

From a historical perspective,
the notion of the spin Hall effect in systems of itinerant spinful charge
carriers was considered first by Dyakonov and Perel \cite{Dyakonov71} 
in the early seventies, and in a more recent paper by Hirsch \cite{Hirsch99}.
In these studies the predicted spin Hall effect is due to spin-orbit
effects influencing scattering processes upon static impurities.
Following the usual terminology of semiconductor physics, this effect
is referred
to as the {\em extrinsic} spin Hall effect since it necessarily requires 
spin-dependent impurity scattering. This is in contrast to the {\em intrinsic}
spin Hall effect which is entirely
due to spin-orbit coupling terms in the single-particle carrier Hamiltonian
and occurs even in the absence of any scattering process. 
The story of the intrinsic spin Hall effect starts in summer of 2003,
when Murakami, Nagaosa, and Zhang \cite{Murakami03}, and,
almost simultaneously, J. Sinova {\it et al.} in a collaboration based in 
Austin (Texas)\cite{Sinova03}
predicted this phenomenon. The work by Sinova {\it et al.}
considers a two-dimensional electron gas being subject to spin-orbit coupling
of the so-called Rashba type, whereas the former paper investigates 
valence-band holes in three-dimensional bulk systems. Shortly
later on, Schliemann and Loss added a study on spin Hall transport of heavy 
holes confined to a quantum well\cite{Schliemann05a}. Only a few months later,
J. Wunderlich {\it et al.} reported on an experimental study of the latter
type of system, confirming the predicted spin Hall effect via optical 
techniques \cite{Wunderlich05}. The paper by Wunderlich {\it et al.} was only
shortly preceded by another experimental report by Kato {\it et al.}
who detected spin Hall transport in n-doped bulk systems, again using an 
optical method \cite{Kato04}. In both experiments, the existence of spin Hall
transport is signaled by spin accumulation at the boundaries of the sample,
which seems to be the easiest method so far to detect this effect.

The theoretical and experimental developments sketched above have generated
a still rapidly growing amount of preprints and (subsequent) journal 
publications, so far mostly theoretical. In this article we give a 
pedagogical overview on important aspects of intrinsic
spin-Hall transport in III-V zinc-blende semiconductors.
Emphasis is put on
the the description of the intrinsic mechanisms of spin Hall transport
induced by spin-orbit coupling,and on the effects of dissipation.
This article also addresses  
researchers who are not particularly specialized in spin phenomena in
semiconductors but are interested in this rapidly developing field.
A brief review on spin-Hall transport was already given in a conference paper
by Murakami \cite{Murakami05a}, and Sinova {\em et al.} recently
provided a short summary of important issues \cite{Sinova05}. 

This article is organized as follows. In section \ref{spin-orbit} we make
a few general comments on spin-orbit coupling in semiconductors and describe
its effective contributions to the band structure of both electron and
hole doped systems. In section \ref{theory} we first make some general remarks
on the notion of spin currents before reviewing the 
particularly rich body of recent theoretical work on spin Hall transport.
Our analysis here includes the two-dimensional electron gas as well
as p-doped bulk systems and quantum wells. Experimental work and proposed
experiments are discussed in section \ref{exp}. We close with conclusions and
an outlook in section \ref{concl}.

\section{Spin-orbit coupling in III-V semiconductors}
\label{spin-orbit}

The coupling between the orbital and the spin degree of freedom of electrons 
is a relativistic effect described by the Dirac equation and its 
nonrelativistic
expansion in powers of the inverse speed of light $c$. 
In second order
one obtains, apart from two spin-independent contributions, the following 
well-known spin-orbit coupling term,
\begin{equation}
{\cal H}_{so}=\frac{1}{2m_{0}c^{2}}\vec s\cdot\left(\nabla V\times\frac{\vec p}{m_{0}}\right)\,,
\label{sogeneral}
\end{equation}
where $m_{0}$ is the bare mass of the electron, $\vec s$, $\vec p$ its spin and
momentum, respectively, and $V$ is some applied external potential. On the 
other hand, the free Dirac equation, $V=0$, has two dispersion branches with
positive and negative energy, 
\begin{equation}
\varepsilon(\vec p)=\pm\sqrt{m_{0}^{2}c^{4}+c^{2}p^{2}}\,,
\end{equation}
which are separated by an energy gap of $2m_{0}c^{2}\approx 1{\rm MeV}$. In particular,
the nonrelativistic expansion of the Dirac equation quoted above can be seen
as a method of systematically including the effects of the negative-energy
solutions on the states of positive energy starting from their nonrelativistic
limit. Moreover, the large energy gap $2m_{0}c^{2}$ appears in the
denominator of the right hand side of Eq.~(\ref{sogeneral}), suppressing the
effects of spin-orbit coupling for weakly bound electrons.
 
On the other hand, the band structure of zinc-blende III-V semiconductors
shows many formal similarities to the situation of free relativistic electrons,
while the relevant energy scales are grossly different
\cite{Zawadzki70,Darnhofer93,Rashba04}. For not too large
doping of such semiconductors, one can concentrate on the band structure around
the $\Gamma$ point. Here one has a parabolic $s$-type conduction band and a
$p$-type valence band consisting of the well-known dispersion branches
for heavy and light holes, and the split-off band. However, the gap between 
conduction and valence band is of order $1{\rm eV}$ or smaller. This heuristic
argument makes plausible that spin-orbit coupling is an important effect in
III-V semiconductors which actually lies at the very heart of the field of
semiconductor spintronics.

In the following we give an overview on effective model Hamiltonians
for conduction-band electrons and valence-band holes in III-V zinc-blende
semiconductors in several spatial
dimensions. These effective expressions can be obtained via the so-called
$\vec k\cdot\vec p$-theory and related methods, as general references we refer to
Refs.~\cite{Kane66,Winkler03,Ivchenko05}. Here we shall just state the results
and discuss their main physical implications.

\subsection{Conduction-band electrons}

Let us first consider three-dimensional bulk systems. For electrons in the
$s$-type conduction band, the contribution to spin-orbit coupling being of
lowest order in the electron momentum $\vec p$ has been derived by
Dresselhaus \cite{Dresselhaus55} and reads
\begin{equation}
{\cal H}_{D}^{bulk}=\frac{\gamma}{\hbar^{3}}\left( 
\sigma^{x}p_{x}\left(p_{y}^{2}-p_{z}^{2}\right)
+\sigma^{y}p_{y}\left(p_{z}^{2}-p_{x}^{2}\right)
+\sigma^{z}p_{z}\left(p_{x}^{2}-p_{y}^{2}\right)\right)\,,
\label{bulkdressel}
\end{equation}
where $\vec\sigma$ is the vector of Pauli matrices describing the electron spin, 
and $\gamma$ is an effective
coupling parameter. This Hamiltonian is trilinear in the momentum
$\vec p$ and invariant under all symmetry operations of the tetrahedral
group $T_{d}$, the point symmetry group of the zinc-blende lattice. 
As a result, the parameter
$\gamma$ is different from zero because the zinc-blende lattice does not
possess an inversion center. Therefore, the Dresselhaus spin-orbit coupling
is due to {\em bulk-inversion asymmetry}.

In a sufficiently narrow quantum well grown along the [001] direction, and 
at sufficiently low temperatures,  
one can approximate the operators $p_{z}$ and $p^{2}_{z}$
by their expectation values 
$\langle p_{z}\rangle\approx 0$, $\langle p^{2}_{z}\rangle=\hbar^{2}\langle k_{z}^{2}\rangle $. Then neglecting terms of order 
$p_{x}^{2}$, $p_{y}^{2}$
leads to a spin-orbit coupling term
linear in the momentum \cite{Dyakonov86,Bastard92},
\begin{equation}
{\cal H}_{D}=\frac{\beta}{\hbar}\left(p_{y}\sigma^{y}-p_{x}\sigma^{x}\right)
\label{dressel}
\end{equation}
with $\beta=\gamma \langle k_{z}^{2}\rangle $. Here $k_{z}$ is the wave number in the lowest 
subband of 
the well. Another important contribution to spin-orbit coupling 
occurs in quantum wells whose confining potential is lacking inversion
symmetry. This contribution due to {\em structure-inversion asymmetry}
is known as the Rashba term \cite{Rashba60,Bychkov84},
\begin{equation}
{\cal H}_{R}=\frac{\alpha}{\hbar}\left(p_{x}\sigma^{y}-p_{y}\sigma^{x}\right),
\label{rashba}
\end{equation}
where  the coupling parameter $\alpha$ is essentially proportional to
the potential gradient across the quantum well. Hence, $\alpha$ is in particular
tunable by an electric gate and can therefore be varied experimentally.

A both theoretically \cite{Lommer88} and experimentally 
\cite{Jusserand92,Jusserand95} well established
value for the Dresselhaus parameter
in GaAs is $\gamma=25{\rm eV}\AA^{3}$. Depending on the width of the quantum well,
this leads to values for $\beta$ being of up to $10^{-11}{\rm eVm}$. Regarding the
Rashba coefficient $\alpha$, values of a few $10^{-11}{\rm eVm}$ can be 
reached in InAs 
\cite{Nitta97,Engels97,Heida98,Hu99,Grundler00,Sato01,Hu03},
whereas in GaAs this quantity is typically an 
order of magnitude smaller \cite{Miller03}. Thus, the 
characteristic energy scales 
\begin{equation}
\varepsilon_{R}=\frac{m\alpha^{2}}{\hbar^{2}}\,,
\end{equation}
\begin{equation}
\varepsilon_{D}=\frac{m\beta^{2}}{\hbar^{2}}
\end{equation}
for Rashba and Dresselhaus coupling, respectively, can be of order
$0.1\dots1.0{\rm meV}$, depending on the effective band mass $m$.

Let us finally briefly discuss the spectrum and eigenstates 
generated by the above spin-orbit coupling terms.
We consider the single-particle Hamiltonian for a two-dimensional electron
system
\begin{equation}
{\cal H}=\frac{\vec p^{2}}{2m}+{\cal H}_{R}+{\cal H}_{D}\,.
\end{equation}
The eigenenergies are given by
\begin{equation}
\varepsilon_{\pm}\left(\vec k\right)
=\frac{\hbar^{2}k^{2}}{2m}
\pm\sqrt{\left(\alpha k_{y}+\beta k_{x}\right)^{2}
+\left(\alpha k_{x}+\beta k_{y}\right)^{2}}
\label{dispersion1}
\end{equation}
with eigenstates
\begin{equation}
\langle\vec r|\vec k,\pm\rangle=
\frac{e^{i\vec k\cdot\vec r}}{\sqrt{A}}\frac{1}{\sqrt{2}}
\left(
\begin{array}{c}
1 \\
\pm e^{i\chi(\vec k)}
\end{array}
\right)
\label{eigenstate}
\end{equation}
where $A$ is the area of the system and
\begin{equation}
\chi(\vec k)=\arg(-\alpha k_{y}-\beta k_{x}+i(\alpha k_{x}+\beta k_{y}))\,.
\label{chi}
\end{equation}
The above spin--orbit coupling terms can be viewed as a momentum-dependent
Zeeman field acting on the electron spin. Consequently, the spin state
of the electron depends on its momentum, as seen in Eq.~(\ref{eigenstate}).
Note that for pure Rashba or Dresselhaus coupling, the dispersions form two
parabolas being shifted horizontally. This is different form a normal Zeeman
field which shifts the dispersion parabolas vertically, i.e. along the
energy axis. The case $\alpha=\pm\beta$ is particular \cite{Schliemann03a,Schliemann03b}. 
Here a new conserved quantity given by $\Sigma:=(\sigma^{x}\mp\sigma^{y})/\sqrt{2}$ arises, 
and the spin state of the electrons becomes independent of the wave vector.
This result for $\alpha=\pm\beta$ is a very general one, it also holds in the presence
of any arbitrary scalar potential, or if interactions between electrons are
included.

\subsection{Valence-band holes}

The valence band of III-V zinc-blende semiconductors is of $p$-type, i.e. it
is predominantly composed out of atomic wave functions with angular momentum
$l=1$. Adding this angular momentum to the electron spin $s=1/2$, 
we find to multiplets with total angular momentum $j=3/2$ and $j=1/2$. The
dublett $j=1/2$ forms essentially an energetically separated 
so-called split-off band and 
will not be considered any further. The multiplet $j=3/2$ consists 
essentially of the so-called heavy and light hole states which are, to a good
degree of approximation, described by Luttinger's Hamiltonian
\cite{Luttinger56},
\begin{equation}
{\cal H}=\frac{1}{2m_{0}}\left(\left(\gamma_{1}+\frac{5}{2}\gamma_{2}\right)
\vec p^{2}-2\gamma_{2}\left(\vec p\cdot\vec S\right)^{2}\right)\,.
\label{Luttinger}
\end{equation}
Here $m_{0}$ is again
the bare electron mass, and $\vec S$ are spin-$3/2$-operators.
The dimensionless Luttinger parameter $\gamma_{1}$ and $\gamma_{2}$ 
describe the valence
band of the specific material with effects of spin-orbit coupling being
included in $\gamma_{2}$. 
The eigenstates of the above Hamiltonian
can be chosen to be eigenstates of the helicity operator
$\lambda=(\vec k\cdot\vec S)/k$, where $\vec k=\vec p/ \hbar$ is the hole wave vector.
The heavy holes correspond to
$\lambda=\pm 3/2$, while the light holes have $\lambda=\pm 1/2$. From the Hamiltonian
(\ref{Luttinger}) one finds the effective band mass for the heavy holes as
\begin{equation}
m_{hh}=\frac{m_{0}}{\gamma_{1}-2\gamma_{2}}
\end{equation}
and for the light holes
\begin{equation}
m_{lh}=\frac{m_{0}}{\gamma_{1}+2\gamma_{2}}\,.
\end{equation}
Well established values for
the Luttinger parameters, among other band structure parameters, can be found 
in the literature \cite{Vurgaftman01}. For example, for GaAs one has
$\gamma_{1}\approx 7.0$ and $\gamma_{2}\approx 2.5$ giving $m_{hh}\approx 0.5m_{0}$ and $m_{lh}\approx 0.08m_{0}$. 

In a bulk
system, the heavy and light hole states are degenerate at the $\Gamma$-point 
$\vec k=0$. This degeneracy is lifted in a quantum well due to size
quantization, and for sufficiently narrow wells and low enough temperatures
one can concentrate on the lower-lying heavy holes. Moreover, if the 
quantized wave vector in the growth direction is large enough, i.e. if the
well is not too wide, the spin of these heavy holes points predominantly
along the growth direction with a projection of $\pm 3/2$. For asymmetric
wells these hole states are subject to a spin-orbit contribution due to
structure-inversion asymmetry analogous to the Rashba term for electrons
in the conduction band. Choosing the growth direction to point along the 
$z$-axis, the resulting effective Hamiltonian has the form 
\cite{Winkler00,Gerchikov92}
\begin{equation}
{\cal H}=\frac{\vec p^{2}}{2m}+i\frac{\tilde\alpha}{2\hbar^{3}}
\left(p_{-}^{3}\sigma_{+}-p_{+}^{3}\sigma_{-}\right)\,,
\label{defham}
\end{equation}
using the notations $p_{\pm}=p_{x}\pm ip_{y}$, 
$\sigma_{\pm}=\sigma^{x}\pm i\sigma^{y}$.
Here the  Pauli matrices operate on the total angular momentum states
with spin projection $\pm 3/2$ along the growth direction; in this sense they
represent a pseudospin degree of freedom rather than a genuine
spin 1/2. In the above equation, $m$
is the heavy-hole mass, and $\tilde\alpha$ is Rashba spin-orbit coupling 
coefficient due to structure inversion asymmetry. 
This Hamiltonian has two dispersion branches given by
\begin{equation}
\varepsilon_{\pm}(k)=\frac{\hbar^{2}k^{2}}{2m}\pm\tilde \alpha k^{3}
\label{dispersion}
\end{equation}
with eigenfunctions
\begin{equation}
\langle\vec r|\vec k,\pm\rangle=
\frac{e^{i\vec k\vec r}}{\sqrt{A}}\frac{1}{\sqrt{2}}
\left(
\begin{array}{c}
1 \\
\mp i\left(k_{x}+ik_{y}\right)^{3}/k^{3}
\end{array}
\right)\,.
\end{equation}
We note that the Rashba parameter $\alpha$ entering the Hamiltonian
(\ref{rashba}) for electrons in a quantum well has a different dimension
than the parameter $\tilde\alpha$ for holes in Eq.~(\ref{defham}). In the latter 
case, the quantity $m\tilde\alpha/ \hbar^{2}$ has dimension of length and can reach
a magnitude of several nanometers in GaAs samples
\cite{Winkler02}.

\section{Spin Hall transport: Theory}
\label{theory}

In this section we summarize important theoretical results on intrinsic
spin Hall effect. We start with some general considerations on spin currents. 

\subsection{Spin currents: General remarks}

A type of current most familiar to physicists is certainly the usual
charge or particle current. The particle density in a many-body system
is described by the operator
\begin{eqnarray}
\rho(\vec r)=\sum_{n}\delta(\vec r-\vec r_{n})
\end{eqnarray}
where the index $n$ labels the particles. This density operator is a function
of time via the time-dependence of the positions $\vec r_{n}$ entering
the argument of the delta-functions. Let this time evolution be generated
by an Hamiltonian of the form
\begin{equation}
{\cal H}=\sum_{n}h(\vec r_{n},\vec p_{n},\vec \sigma_{n})+V_{int}\,.
\end{equation}
Here the term $V_{int}$ describes interaction among the particles and depends
only on their spatial coordinates,
and the single-particle Hamiltonian $h$ reads
\begin{equation}
h(\vec r,\vec p_,\vec \sigma)=\frac{\vec p^{2}}{2m}+\gamma_{ij}p_{i}\sigma^{j}+V(\vec r)\,.
\end{equation}
Summation over repeated cartesian indices is understood, and the matrix
$\gamma$ parameterizes spin-orbit coupling of the Rashba and Dresselhaus type
for electrons in a quantum well.  Finally, the static potential $V(\vec r)$ 
describes, e.g., static impurities.
Now, starting from the Heisenberg equation
of motion,
\begin{equation}
\frac{d}{dt}\rho=\frac{i}{\hbar}\left[{\cal H},\rho\right]
\end{equation}
and performing some elementary algebraic manipulations,
one derives the well-known continuity equation for the particle current,
\begin{equation}
\frac{d}{dt}\rho+\nabla\cdot\vec j=0
\end{equation}
with the current density operator
\begin{equation}
\vec j(\vec r)=\frac{1}{2}\sum_{n}\left\{\vec v(\vec p_{n},\vec\sigma_{n}),
\delta(\vec r-\vec r_{n})\right\}\,.
\end{equation}
Here $\{A,B\}=AB+BA$ denotes the anticommutator of two operators,
and the velocity operator $\vec v$ is given by
\begin{equation}
\vec v(\vec p_{n},\vec\sigma_{n})=\frac{i}{\hbar}\left[{\cal H},\vec r_{n}\right]
\end{equation}
for each particle $n$.
Note that this operator is in general spin-dependent if spin-orbit coupling is
present.

Let us now consider a general observable described by a hermitian
single-particle operator $A$ which can be a function of position, momentum,
and spin: $A=A(\vec r,\vec p,\vec\sigma)$. The density operator
corresponding to this physical quantity $A$ is naturally defined as
\begin{equation}
\rho_{A}(\vec r)
=\frac{1}{2}\sum_{n}\left\{A(\vec r_{n},\vec p_{n},\vec\sigma_{n}),\delta(\vec r-\vec r_{n})\right\}\,,
\end{equation}
where the symmetrization ensures hermiticity. Now proceeding as above one finds
\begin{equation}
\frac{d}{dt}\rho_{A}+\nabla\cdot\vec j_{A}=s_{A}\,,
\end{equation}
where the current density operator $\vec j_{A}$ is given by
\begin{equation}
\vec j_{A}(\vec r)=\frac{1}{4}\sum_{n}\left\{A(\vec r_{n},\vec p_{n},\vec\sigma_{n}),
\left\{\vec v(\vec p_{n},\vec\sigma_{n}),
\delta(\vec r-\vec r_{n})\right\}\right\}\,,
\end{equation}
and the additional source term on the right-hand-side  reads
\begin{equation}
s_{A}(\vec r)
=\frac{1}{2}\sum_{n}\left\{
\frac{i}{\hbar}\left[{\cal H},A(\vec r_{n},\vec p_{n},\vec\sigma_{n})\right],
\delta(\vec r-\vec r_{n})\right\}\,.
\end{equation}
Thus, we only obtain the usual from of the continuity equation if 
the observable $A$ commutes with the Hamiltonian ${\cal H}$, which is of
course just a restatement of Noether's theorem. For the case electron spin
components as observables, the corresponding spin-current densities
are given by 
\begin{equation}
\vec j_{i}(\vec r)=\frac{1}{4}\sum_{n}\left\{\frac{\hbar}{2}\sigma^{i}_{n},
\left\{\vec v(\vec p_{n},\vec\sigma_{n}),
\delta(\vec r-\vec r_{n})\right\}\right\}\,,
\label{defspincurr}
\end{equation}
and the source terms are due to spin-orbit coupling being present in the
single-particle Hamiltonian. These source terms reflect the fact that 
magnetization, i.e. the density of magnetic moments, can be altered by two
ways: by spatially moving spinful particles, or by manipulating their spin 
state. The latter process is described by the source terms. For instance,
in the case of noninteracting electrons in a quantum well with 
spin-orbit coupling of the Rashba and Dresselhaus type, the source 
terms can be expressed via the components of the spin-current densities itself
\cite{Erlingsson05a,Burkov04}, 
\begin{eqnarray}
\frac{d}{dt} \rho_{x}
+\nabla\cdot\vec j_{x}
 & = & \frac{2m\alpha}{\hbar^{2}}j^{x}_{z}+\frac{2m\beta}{\hbar^{2}}j^{y}_{z}\,,
\label{spincont1}\\
\frac{d}{dt} \rho_{y}
+\nabla\cdot\vec j_{y}
 & = & \frac{2m\alpha}{\hbar^{2}}j^{y}_{z}+\frac{2m\beta}{\hbar^{2}}j^{x}_{z}\,,
\label{spincon21}\\
\frac{d}{dt} \rho_{z}
+\nabla\cdot\vec j_{z}
 & = & -\frac{2m\alpha}{\hbar^{2}}\left(j^{x}_{x}+j_{y}^{y}\right)
-\frac{2m\beta}{\hbar^{2}}\left(j^{x}_{y}+j^{y}_{x}\right).
\label{spincont3}
\end{eqnarray}
In summary, the definition of the spin current density as given in 
Eq.~(\ref{defspincurr}) is the straightforward generalization of the 
usual particle current and widely used in the literature. As seen above,
this spin current density is, however, not conserved, i.e. it does not
fulfill a simple continuity equation. This fact might or might not be seen as 
a shortcoming of the above definition. Another peculiarity of this type
of current operator was pointed out by Rashba \cite{Rashba03} who found
that, for the situation of an asymmetric quantum well, the current densities
with in-plane spin components, $\vec j_{x}$, $\vec j_{y}$, can have nonzero
expectation values even in the absence of an electric field, i.e. in thermal
equilibrium. Using periodic boundary conditions and considering an infinite 
disorder-free system of non-interacting electrons 
at zero temperature and positive Fermi energy, 
the full result for the case of both Rashba and
Dresselhaus coupling reads \cite{Erlingsson05a}
\begin{eqnarray}
\langle j^{x}_{x}\rangle=-\langle j^{y}_{y}\rangle & = & 
\frac{\beta}{6\pi}\left(\frac{m}{\hbar^{2}}\right)^{2}
\left(\alpha^{2}-\beta^{2}\right)\,,
\label{equi1}\\
\langle j^{x}_{y}\rangle=-\langle j^{y}_{x}\rangle & = & 
\frac{\alpha}{6\pi}\left(\frac{m}{\hbar^{2}}\right)^{2}
\left(\alpha^{2}-\beta^{2}\right)\,.
\label{equi2}
\end{eqnarray}
Note that these equilibrium spin currents vanish in the case $\alpha=\pm\beta$ due
to the additional conserved spin operator arising at this point
\cite{Schliemann03a}. 
The findings shown in Eqs.~(\ref{equi1}),(\ref{equi2}), 
however, certainly depend on the boundary conditions used and
are altered in a more realistic description of finite systems
\cite{Kiselev05}.

In the recent literature, there are several proposals and discussions on 
alternative forms of spin currents which possibly fulfill proper
continuity equations 
\cite{Murakami04a,Rashba04a,Jin05,Zhang05,Li05,Sugimoto05}. However, 
these issues do not seem to be settled yet. Therefore, in the following we
shall concentrate on spin current densities as defined in 
Eq.~(\ref{defspincurr}).

\subsection{Conduction-band electrons in two dimensions}

We now discuss spin Hall transport of conduction-band electrons in 
III-V semiconductor quantum wells. 
As a great simplification used in almost
the entire theoretical work so far, we will consider non-interacting 
electrons. Thus, the system is described by the single-particle Hamiltonian
\begin{equation}
{\cal H}=\frac{\vec p^{2}}{2m}+\frac{\alpha}{\hbar}\left(p_{x}\sigma^{y}-p_{y}\sigma^{x}\right)
+\frac{\beta}{\hbar}\left(p_{y}\sigma^{y}-p_{x}\sigma^{x}\right)\,,
\label{singpartham}
\end{equation}
and instead of the many-body spin current density operators
(\ref{defspincurr}) we can use the single-particle operator
\begin{eqnarray}
\vec j_{z} & = & \frac{\hbar}{4}\left(\sigma^{z}\vec v+\vec v\sigma^{z}\right)\\
 & = & \frac{\vec p}{m}\frac{\hbar}{2}\sigma^{z}\,,
\end{eqnarray}
where we have concentrated on the spin component along the growth
direction of the quantum well (chosen as z-axis) and used the anticommutativity
of Pauli matrices. To account for effects of disorder and confining 
boundaries of the system, appropriate
potentials should be added to the Hamiltonian (\ref{singpartham}),
as we will discuss in detail below.
 
The linear response of this spin current to an electric
field applied in the plane of the two-dimensional electron gas can be
evaluated via the usual Kubo formula \cite{Mahan00}. For the
off-diagonal (or Hall) components of the response tensor one has
\cite{Sinova03,Schliemann04}
\begin{equation}
\sigma^{S,z}_{xy}(\omega)=\frac{e}{A(\omega+i\eta)}
\int_{0}^{\infty}e^{i(\omega+i\eta)t}\sum_{\vec k,\mu}f(\varepsilon_{\mu}(\vec k))
\langle\vec k,\mu|[j^{x}_{z}(t),v_{y}(0)]|\vec k,\mu\rangle\,.
\label{generalKubo}
\end{equation}
Here $A$ is the volume of the system, $e$ is the elementary charge, and
$f(\varepsilon_{\mu}(\vec k))$ is the Fermi distribution function 
for  energy $\varepsilon_{\mu}(\vec k)$ at wave vector $\vec k$ in the
dispersion branch $\mu=\pm$ as given in Eq.~(\ref{eigenstate}).
The above quantity describes the linear response in terms of a spin current
to a perpendicular electric field of
frequency $\omega$. In the commutator on the right-hand side the 
time-dependent spin current operator in the Heisenberg picture enters,
\begin{equation}
\vec j_{z} (t)=e^{i{\cal H}t/\hbar}\vec j_{z}e^{-i{\cal H}t/\hbar}\,.
\end{equation}
Moreover, the right-hand side of Eq.~(\ref{generalKubo}) has to be understood
in the limit of vanishing imaginary part $\eta>0$ in the frequency argument.
This imaginary part in the frequency reflects the
fact that the external electric field is assumed to be switched on 
adiabatically starting from the infinite past of the system, and it also
ensures causality properties of the retarded Green's function occurring in
Eq.~(\ref{generalKubo}). In general, and as we will discuss in more detail 
below, the limiting process $\eta\to 0$ does not commute with other
limits, and, in particular, the dc-limit $\omega\to 0$ has to be taken with 
care \cite{Mahan00}. In the presence of random impurity scattering,
the retarded two-body Green's function in Eq.~(\ref{generalKubo}) will 
generically have
a frequency argument with positive imaginary part \cite{Mahan00}. 
In this case the limit 
$\eta\to 0$ is unproblematic, and the imaginary part of the
frequency argument is just due to impurity scattering and/or other
(many-body) effects. Generically, the imaginary part $\eta>0$ corresponds to a 
finite carrier quasiparticle lifetime. 

Let us now for simplicity consider the case of Rashba spin-orbit coupling only
and assume the electron density to be large enough such that the Fermi
energy is positive (which is usually the case in realistic samples).
Neglecting all possible disorder effects and concentrating on the case of 
zero temperature, the Kubo formula
(\ref{generalKubo}) can be evaluated straightforwardly,  giving 
a spin Hall conductivity at zero frequency of
\begin{equation}
\sigma^{S,z}_{xy}(0)=-\sigma^{S,z}_{yx}(0)=\frac{e}{8\pi}\,.
\end{equation}
This result was obtained first by Sinova {\em et al.} \cite{Sinova03}. It is
remarkable in the sense that the value of the spin Hall conductivity does
not depend on the Rashba parameter $\alpha$. In particular, even in the limit of
vanishing spin-orbit coupling, the above result still predicts a finite
spin Hall current. However, no effects of disorder in the system have been 
included so far, and the infinitesimal parameter $\eta$ has been put
to zero right away. Let us now take into account disorder effects by 
replacing $\eta$ with a phenomenological relaxation rate $1/\tau$. Here we find
\cite{Schliemann04}
\begin{equation}
\sigma^{S,z}_{xy}(0)=-\sigma^{S,z}_{yx}(0)
=\frac{e}{8\pi}
-\frac{e}{32\pi}\frac{\hbar/\tau}{\varepsilon_{R}}\tan^{-1}\left(4\frac{\varepsilon_{R}}{\hbar/\tau}
\left(1+8\frac{\varepsilon_{R}\varepsilon_{f}}
{(\hbar/\tau)^{2}}\right)^{-1}\right)\,.
\label{SHtau}
\end{equation}
The first term is still the universal expression found in Ref.~\cite{Sinova03},
whereas in the second contribution three energy scales enter: The Fermi
energy $\varepsilon_{f}$, the Rashba energy $\varepsilon_{R}$, and the energy scale of the 
scattering by disorder potentials $\hbar/ \tau$. Clearly, if the latter quantity 
dominates over the Rashba coupling, $\hbar/ \tau\gg\varepsilon_{R}$, the second term in
Eq.~(\ref {SHtau}) cancels the first one, and the spin Hall conductivity
is indeed suppressed by disorder. Analogous results can be found if
both the Rashba and the Dresselhaus coupling are included
\cite{Sinitsyn04}. In this case, the above approach also yields nonvanishing
{\em longitudinal} spin conductivities \cite{Sinitsyn04}.

Thus, we arrive at an apparently physically 
satisfactory picture. However, it turns out to be qualitatively incorrect
for the following reason: When replacing the infinitesimal
parameter $\eta$ in Eq.~(\ref{generalKubo}) with the inverse of a 
phenomenological relaxation time $\tau$, one neglects certain contributions in
a systematic perturbational expansion with respect to the disorder potentials.
These contributions are known as vertex corrections. As is was shown first
by Inoue, Bauer, and Molenkamp \cite{Inoue04}, the full dissipative 
contribution to the spin-Hall conductivity including the vertex corrections
{\em exactly cancels the universal value, independently of the
strength of the disorder potentials and the Rashba coupling}. This result
was obtained for an infinitely large system and
in lowest perturbational order with respect to the disorder
potentials which were modeled by delta-functions. Subsequently, this finding 
was reproduced an generalized by several other authors
\cite{Mishchenko04,Khaetskii04,Raimondi05,Chalaev05,Dimitrova05,Liu06,Grimaldi06}
using different theoretical methods. The
conclusions from these investigations can be summarized as follows:
The spin-Hall conductivity for spin polarization along
the growth direction in an infinite two-dimensional system with spin-orbit
coupling of the Rashba and Dresselhaus type strictly vanishes
in the presence of any spin-independent mechanism 
with forces, via spin-orbit coupling, the electron spins to relax
to a constant value. This result is independent of any perturbational
expansion with respect to disorder terms and holds also both 
at finite temperature and in the presence of interactions among the electrons.
However, it does in general not hold in the presence of magnetic fields
or other spin-dependent contributions in the Hamiltonian.

The proof of this very general statement was worked out by
Chalaev and Loss \cite{Chalaev05}, and by Dimitrova \cite{Dimitrova05};
a preliminary version can also be found in 
Ref.~\cite{Erlingsson05a}. For the case of both 
Rashba and Dresselhaus spin-orbit coupling, the argument is as follows: 
To analyze the electron spin dynamics in an infinite homogeneous system, it
is sufficient to consider just the spin operator of a single electron.
From the Heisenberg equation of motion we find
\begin{eqnarray}
\frac{d}{dt}\sigma^{x} & = & \frac{4m\alpha}{\hbar^{3}}j_{z}^{x}+\frac{4m\beta}{\hbar^{3}}j_{z}^{y}\,,
\label{timderv1}\\
\frac{d}{dt}\sigma^{y} & = & \frac{4m\beta}{\hbar^{3}}j_{z}^{x}+\frac{4m\alpha}{\hbar^{3}}j_{z}^{y}\,.
\label{timderv2}
\end{eqnarray}
These relations hold also in the presence of any arbitrary spin-independent
potential or interaction term in the Hamiltonian. The key observation is
that the time derivatives of the spin components can be expressed as
linear combinations of the spin current operators itself. This result crucially
relies on the fact that the spin-orbit coupling is linear in the electron 
momentum. Moreover, in the presence of spin-orbit coupling disorder effects
can generally be expected to make the electron spins relax to constant values.
Thus, in a stationary state, the expectation values of the left-hand sides
of Eqs.~(\ref{timderv1}), (\ref{timderv2}) should vanish. Now it follows
immediately that the expectation values of the spin current components
$j_{z}^{x}$, $j_{z}^{y}$ must also vanish provided $\alpha\neq\pm\beta$. In the case $\alpha=\pm\beta$, however,
spin Hall transport is generally absent due to the additional conserved
spin quantity which occurs here \cite{Sinitsyn04,Shen04a,Schliemann03a}.

The central argument of the above proof is closely related to studies
by Erlingsson, Schliemann and Loss \cite{Erlingsson05a}, and by
Dimitrova \cite{Dimitrova05} on the relationship between spin currents
and magnetic susceptibilities. For further developments in this direction see
also Ref.~\cite{Bernevig05a}. 

Another interesting observation regarding spin Hall transport in n-doped
quantum wells was made by Rashba \cite{Rashba04b} who considered Rashba 
spin-orbit coupling in the presence of a magnetic field coupling to the
orbital degrees of freedom only, neglecting the Zeeman coupling to the
electron spin. In this model, the spin Hall conductivity vanishes 
in the limit of vanishing magnetic field even in the
absence of disorder, an effect closely related to the abovementioned 
vertex corrections \cite{Rashba04b}. Thus, the case of zero magnetic field 
(coupling to the orbital degrees of freedom only) is 
different from the limit of vanishing field. This result certainly relies 
on the assumption of an infinite system, since the orbital effects
of a magnetic field should be small if the typical cyclotron radius is large 
compared to the system size.

The case of a magnetic field coupling both to the orbital and spin degree of
freedom of electrons was investigated by Shen {\em et al} 
\cite{Shen04b,Shen05}. Due to the coupling to the spin this
situation is not covered by the above general argument.
Shen {\em et al} find a resonant behavior of the
spin Hall conductance as a function of the magnetic field
when a degeneracy of Landau levels occurs at Fermi level. These studies, 
however, do not include disorder effects so far \cite{Shen04b,Shen05}.

Coming back the case zero external magnetic field,
Adagideli and Bauer chose yet another approach to the problem by 
considering the electron acceleration in the presence of Rashba coupling
\cite{Adagideli05},
\begin{equation}
\frac{d^{2}}{dt^{2}}\vec r=\frac{4m\alpha^{2}}{\hbar^{4}}\vec j_{z}\times\vec e_{z}
-\frac{1}{m}\left(e\vec E+\nabla V \right)\,,
\end{equation}
where $\vec e_{z}$ is the unit vector in the $z$-direction, $\vec E$ is the
in-plane electric field, and $V$ is the disorder potential. Since the 
acceleration should relax to zero in a disordered system, the spin Hall
current vanishes if the effects of the electric field and the disorder
potential cancel on average. As the authors show, this is indeed the case
in the bulk of the system, but not necessarily at its edges. This observation
gives rise to the notion of {\em spin Hall edges} \cite{Adagideli05}.

The above general result on the absence of spin Hall transport in 
infinite systems with spin-orbit coupling being linear in the electron
momentum is the outcome of an intense theoretical discussion in the last
about two years. It was also confirmed by numerical studies carried out by
Nomura {\em et al.} \cite{Nomura05a}. 
These authors performed a careful numerical evaluation
of the Kubo formula for the spin Hall conductivity in the presence of
Rashba coupling and delta-function type impurity potentials.

We stress again that the above general conclusion rules out spin Hall transport
only in the limit of an infinite system. In fact, several numerical 
investigations on finite systems have appeared recently
\cite{Hankiewicz04,ShengL05a,LiJ05a,Nikolic05a,Nikolic05b,Hankiewicz05,ShengD05a,Moca05,Nikolic06}.
In these studies, the underlying semiconductor structure are described by 
tight-binding hopping models of finite-size
coupled to semi-infinite leads. These hopping
Hamiltonians also include local disorder potentials and discrete
versions of the spin-orbit contributions in an n-doped quantum well.
Transport quantities are then evaluated using the well-established
Landauer-B\"uttiker approach combined with a Green's function treatment of the
semi-infinite leads\cite{Datta95}. In summary, it still remains an
interesting and unsettled question, whether intrinsic spin Hall transport 
can be experimentally observed in mesoscopic systems as 
studied in the above references.

\subsection{Spin Hall transport of holes}

We now analyze intrinsic spin Hall transport in p-doped III-V semiconductors
and start with the case of a three-dimensional bulk system pioneered by
Murakami, Nagaosa, and Zhang \cite{Murakami03}.

\subsubsection{Three-dimensional bulk case}

We consider valence-band heavy and light holes governed by the
Hamiltonian (\ref{Luttinger}) and the conventionally defined spin current
operator as described before. In this case a nonzero spin conductivity occurs
if the direction of the spin current, its spin polarization, and the
driving electric field are mutually orthogonal. For definiteness, let us
assume the spin polarization to point alone the $z$-axis with the electric 
field being in the $xy$-plane. From the Kubo formula
(\ref{generalKubo}), the zero-frequency spin Hall conductivity can be 
evaluated as \cite{Schliemann04,Murakami04a,Culcer04}
\begin{equation}
\sigma_{xy}^{S,z} (0)=\frac{e}{4\pi^{2}}\frac{\gamma_{1}+2\gamma_{2}}{\gamma_{2}}\left(k_{f}^{h}-k_{f}^{l}
-\int_{k_{f}^{l}}^{k_{f}^{h}}dk
\frac{1}{1
+\left(\frac{2}{\hbar/\tau}\frac{\hbar^{2}}{m}\gamma_{2} k^{2}\right)^{2}}
\right)\,,
\label{holesbulkcond}
\end{equation}
where we have again assumed an infinite system at zero temperature,
and 
\begin{equation}
k_{f}^{h/l}=\sqrt{\frac{2m}{\hbar^{2}}\varepsilon_{f}
\frac{1}{\gamma_{1}\mp2\gamma_{2}}}
\end{equation}
are the Fermi wave numbers for heavy and light holes,
respectively. Similarly to the approach leading to Eq.~(\ref{SHtau})
for conduction-band electrons, disorder effects are taken into account via
an effective relaxation time $\tau$. In the case here this approach is
justified because vertex corrections can be shown to be absent for
scattering potentials described by delta-functions\cite{Murakami04b}.
Thus, for this type of disorder, the result (\ref{holesbulkcond}) is exact
in lowest order Born approximation \cite{Mahan00}. The absence of vertex
corrections in this case crucially relies on the fact that 
the underlying Hamiltonian (\ref{Luttinger}) is not linear but of second
order in the components of the momentum \cite{Murakami04b,Bernevig05b}.
The case of impurity potentials of longer spatial range was investigated
in Ref.~\cite{Liu05a}.

The remaining integral in Eq.~(\ref{holesbulkcond})
is elementary leading to a rather tedious
expression which shall not be given here.
However, we
see that the value of the above integral is governed by the ratio of
energy scale of the impurity scattering
$\hbar/\tau$ and the ``spin-orbit energy''
\begin{equation}
\varepsilon_{so}:=\hbar^{2}\gamma_{2}(k_{f}^{0})^{2}/m=2\varepsilon_{f}\gamma_{2}/\gamma_{1}, 
\end{equation}
since 
\begin{equation}
k_{F}^{0}=\sqrt{2m\varepsilon_{f}/\gamma_{1}\hbar^{2}}
\end{equation}
is a typical
wave number in the integration interval \cite{Schliemann04}. If 
$\hbar/\tau\gg \varepsilon_{so}$ the spin Hall
conductivity vanishes as
\begin{eqnarray}
\sigma_{xy}^{S,z} (0) & = & 
\frac{e}{\pi^{2}}4k_{f}^{0}
\left(\frac{\varepsilon_{so}}{\hbar/\tau}\right)^{2}
\frac{\gamma_{2}}{\gamma_{1}}\nonumber\\
 & & +{\cal O}\left(\left(\frac{\varepsilon_{so}}{\hbar/\tau}\right)^{4}
,\left(\frac{\varepsilon_{so}}{\hbar/\tau}\right)^{2}
\left(\frac{\gamma_{2}}{\gamma_{1}}\right)^{2}\right)
\label{expansion1}
\end{eqnarray}
where we have also assumed that the ratio $\gamma_{2}/\gamma_{1}$ is small
as it is usually the case \cite{Vurgaftman01}.
In the opposite case $\hbar/\tau\ll \varepsilon_{so}$ one finds
\begin{eqnarray}
\sigma^{S,z}_{xy}(0)
 & = & \frac{e}{4\pi^{2}}\frac{\gamma_{1}+2\gamma_{2}}{\gamma_{2}}
\left[k_{f}^{h}-k_{f}^{l}+\frac{\left(k_{f}^{0}\right)^{4}}{12}
\left(\left(\frac{1}{k_{f}^{h}}\right)^{3}
-\left(\frac{1}{k_{f}^{l}}\right)^{3}\right)
\left(\frac{\hbar/\tau}{\varepsilon_{so}}\right)^{2}\right]\nonumber\\
 & & +{\cal O}\left(\left(\frac{\hbar/\tau}{\varepsilon_{so}}\right)^{4}\right)
\label{expansion2}
\end{eqnarray}
Here the the contribution in leading order is the result obtained in
Refs.~\cite{Murakami04a,Culcer04} for a disorder-free system
(up to some definitorial prefactor 
\cite{Schliemann04}). The expression given originally in 
Ref.~\cite{Murakami03}, however, 
differs somewhat from the above one due some approximation
employed there \cite{Murakami03}.
Ref.~\cite{Bernevig04a} contains calculations of the spin Hall
conductivity in band structure models more general than Eq.~(\ref{Luttinger}).
Numerical results based on {\em ab initio} band structure calculations 
were presented in Ref.~\cite{Guo05}.
A further numerical study of spin Hall transport within the Hamiltonian
(\ref{Luttinger}) in the presence of disorder was performed in 
Ref.~\cite{Chen05a}.

In summary, spin Hall transport in p-doped
bulk III-V semicondcutors is robust against disorder of not too large
strength, but naturally breaks down if impurity effects are overwhelming
the spin-orbit coupling \cite{Schliemann04}. 

\subsubsection{Heavy holes in a quantum well}

Let us now turn to the case of spin Hall transport of heavy holes in
p-doped quantum wells which was studied first by Schliemann and Loss
\cite{Schliemann05a}. An experimental observation of spin Hall effect
in such a system was recently reported by Wunderlich {\em et al.}
\cite{Wunderlich05}.

We consider the Hamiltonian (\ref{defham}) and
a spin current
 \begin{equation}
\vec j_{z}=\frac{\vec p}{m}\frac{3\hbar}{2}\sigma^{z}
\end{equation}
for heavy holes with spin $\pm 3/2$ polarized along the growth direction of the 
well.
Proceeding as above, one finds for the zero-frequency
spin Hall conductivity \cite{Schliemann05a}
\begin{equation}
\sigma^{S,z}_{xy}(0)=-\sigma^{S,z}_{yx}(0)
=\frac{e}{\pi}\frac{9}{4}\frac{\hbar^{2}\tilde\alpha}{m}\int_{k_{f}^{+}}^{k_{f}^{-}}dk
\frac{k^{4}}{\left(\hbar/\tau\right)^{2}+
\left(2\tilde\alpha k^{3}\right)^{2}}\,,
\label{fullspinHall}
\end{equation}
where $\tau$ is again the momentum relaxation time. Similar to the 
three-dimensional bulk
case, vertex corrections to the spin-Hall conductivity turn out to
be zero for delta-function shaped scatterers \cite{Bernevig05b}, justifying
the above approach. 
This result is again
due to the fact that the spin-orbit coupling is not linear but of higher 
order in the particle momentum.

The Fermi wave numbers $k_{f}^{\pm}$ entering Eq.~(\ref{fullspinHall}) refer to the
two dispersion branches (\ref{dispersion}) and can be
expressed in terms of the particle density
\begin{equation}
n=\frac{1}{4\pi}
\left(\left(k_{f}^{+}\right)^{2}+\left(k_{f}^{-}\right)^{2}\right)
\label{density}
\end{equation}
as\cite{Schliemann05a}
\begin{eqnarray}
k_{f}^{\pm} & = & \mp\frac{1}{2}\frac{\hbar^{2}}{2m\tilde\alpha}
\left(1-\sqrt{1-\left(\frac{2m\tilde\alpha}{\hbar^{2}}\right)^{2}4\pi n}\right)
\nonumber\\
& & +\sqrt{-\frac{1}{2}\left(\frac{\hbar^{2}}{2m\tilde\alpha}\right)^{2}
\left(1-\sqrt{1-\left(\frac{2m\tilde\alpha}{\hbar^{2}}\right)^{2}4\pi n}\right)
+3\pi n}\,.
\label{kfpm}
\end{eqnarray}
Moreover, the {\em longitudinal} spin conductivities $\sigma^{S,z}_{xx}$,
$\sigma^{S,z}_{yy}$ turn out out be identically zero. 

The remaining integral in the above expression (\ref{fullspinHall})
is elementary; however, 
it leads to a rather cumbersome expression which shall again not be given here.
Analogously to the previous case, the energy scale of impurity scattering 
$\hbar/\tau$ has to be compared with the ``Rashba energy'' 
$\tilde\varepsilon_{R}=\tilde\alpha (k_{f}^{0})^{3}$, where 
$k_{f}^{0}=\sqrt{2m\varepsilon_{f}/\hbar^{2}}$ is the Fermi wave number
for vanishing spin-orbit coupling, which is a typical value
for $k$ in the integration in Eq.~(\ref{fullspinHall}).
If the impurity scattering dominates over the Rashba coupling,
$\hbar/\tau\gg\tilde\varepsilon_{R}$, the spin Hall conductivity vanishes
with the leading order correction given by
\begin{equation}
\sigma^{S,z}_{xy}(0)=\frac{e}{\pi}\frac{9}{20}\frac{\tilde\alpha}{m}\tau^{2}
\left((k_{f}^{-})^{5}-(k_{f}^{+})^{5}\right)
+{\cal O}\left(\left(\frac{\tilde\varepsilon_{R}}{\hbar/\tau}\right)^{4}\right)
\,,
\label{spinHall1}
\end{equation}
where the Fermi wave numbers are given by Eq.~(\ref{kfpm}). In the opposite 
case $\tilde\varepsilon_{R}\gg\hbar/\tau$, the leading contribution to the 
spin Hall conductivity reads
\begin{equation}
\sigma^{S,z}_{xy}(0)=\frac{e}{\pi}\frac{9}{16}
\frac{\hbar^{2}}{m\tilde\alpha}
\left(\frac{1}{k_{f}^{+}}-\frac{1}{k_{f}^{-}}\right)
+{\cal O}\left(\left(\frac{\hbar/\tau}{\tilde\varepsilon_{R}}\right)^{4}\right)\,.
\label{spinHall2}
\end{equation}
Note that this result for the spin Hall conductivity depends only on the
length scale $m\tilde\alpha/\hbar^{2}$ of the Rashba coupling and the total hole
density $n$, but not separately on quantities like the Fermi energy and
the effective heavy hole mass. If $m\tilde\alpha/\hbar^{2}$ is small against
the inverse square root of the total hole density (but still fulfilling
$\tilde\varepsilon_{R}\gg\hbar/\tau$), the spin Hall conductivity
approaches a value of 
\begin{equation}
\sigma^{S,z}_{xy}=9\frac{e}{8\pi}\,. 
\label{lowdis}
\end{equation}
This is the case if $\hbar/\tau\ll\tilde\varepsilon_{R}\ll\varepsilon_{f}$. 
This above value should be  
compared with the universal value of $e/8\pi$ found originally
in Ref.\cite{Sinova03} for electrons in a fully clean asymmetric quantum well.
In this sense the hole spin Hall conductivity is enhanced by a factor of $9$
compared to the naive result for electrons, which is partially due to the 
larger angular momentum of the heavy holes. 

Zarea and Ulloa studied the above system in the clean limit but with a 
perpendicular homogeneous magnetic field coupling to the
orbital degrees of freedom of the holes but not to their spin
\cite{Zarea06}, an investigation analogous to the one 
by Rashba on n-doped quantum wells already mentioned \cite{Rashba04b}.
Again it is found that, for an infinite system, the case of zero
magnetic field and the limit of vanishing magnetic field do not 
coincide \cite{Zarea06}. The details of this effect, however, seem to be
somewhat different from the observations made in Ref.~\cite{Rashba04b} and need
further study. In any case, influence of a magnetic field coupling to the
orbital degrees of freedom only should only be appreciable if the field is
strong enough to produce  typical cyclotron radii being of order of the 
system size or smaller. Therefore, arbitrarily small fields cannot be
expected to have an effect in real experiments.
Another study of spin Hall transport in the presence of a perpendicular
magnetic field was performed in Ref.~\cite{Ma05} where a more general
band structure Hamiltonian was used \cite{Bernevig05b}

As seen above, spin Hall transport of heavy holes in a quantum well is
robust against disorder effects, differently from the situation for
electrons in n-doped wells. This effect is due to the different functional 
form of the effective spin-orbit coupling and was also
confirmed numerically by Nomura {\em et al.} who performed a careful
comparison between these two systems \cite{Nomura05a}. In a subsequent
study, the edge- spin accumulation caused by the spin current was
investigated numerically \cite{Nomura05b}. Further investigations
of disorder effects can be found in Ref.~\cite{Liu05} where an approach
based on nonequilibrium Green's functions was used.

Moreover, several groups have studied numerically tight-binding models
for two-dimensional hole systems coupled to semi-infinite leads. 
\cite{Hankiewicz05,Wu05,Chen05}. These investigations are analogous in 
spirit and technical approach to the numerical work on models for n-doped
systems mentioned earlier
\cite{Hankiewicz04,ShengL05a,LiJ05a,Nikolic05a,Nikolic05b,ShengD05a,Moca05,Nikolic06}. In particular, the numerical results on heavy-hole systems
\cite{Hankiewicz05,Chen05} in the limit of low disorder
confirm quantitatively the the enhanced spin conductivity
(\ref{lowdis}) obtained analytically in Ref.~\cite{Schliemann05a}.
Finally, many of the abovementioned mainly numerical studies on p-doped
quantum wells were inspired
by the experiments by Wunderlich {\em et al.} which we will discuss in section
\ref{exp}. 

\subsection{Spin Hall effect in other systems}

Intrinsic mechanisms of spin Hall transport in n-doped bulk III-V 
semiconductors where investigated by Bernevig and Zhang 
\cite{Bernevig05c,Bernevig05d}. Here the leading contribution to spin-orbit
coupling is given by the bulk Dresselhaus term (\ref{bulkdressel}) being
of third order in the electron momentum. On the other hand, Engel, Halperin, 
and Rashba have studied extrinsic spin Hall effect in such systems
\cite{Engel05}. We will discuss these issues in section \ref{exp} in the 
context of the experimental results by Kato {\em et al.} \cite{Kato04}.

Spin Hall effect in graphene, i.e. single planes of graphite, was investigated
by Kane and Mele \cite{Kane05}. Using a theoretical picture similar
to the edge-state theory of the charge quantum Hall effect, 
these authors propose a spin current at the 
edges of a graphene sheet. Work following these theoretical predictions
include Refs.~\cite{Sheng05,Yang06}.
Finally, Shchelushkin and Brataas considered spin Hall transport in
normal metals due to extrinsic mechanisms
\cite{Shchelushkin05}.

\section{Detection of spin Hall transport: 
Experiments and Proposals}
\label{exp}

We now turn to experimental investigations of spin Hall effect 
in semiconductors.
The studies already carried out and many of the experimental proposals
found in the literature use the spin accumulation caused by the spin current
for detecting spin Hall transport. Using a simple spin diffusion model, this
spin accumulation is expected to decay towards the bulk of the sample
on a length scale given by the spin diffusion length \cite{Zhang00}. 
The latter quantity is determined by the semiconductor  material, but
possibly also by further details of the sample and the experimental setup
\cite{Tse05}. Further recent theoretical studies on spin accumulation
caused by the spin Hall effect include 
\cite{Nomura05b,Ma04,Hu04,Usaj05,Malshukov05,Reynoso05}.

Kato {\em et al.} have studied spin Hall  transport in n-doped bulk
epilayers of GaAs and InGaAs with a thickness of $2\mu{\rm m}$ and $500{\rm nm}$,
respectively \cite{Kato04}. The electron density in both samples 
was $n=3\cdot10^{16}{\rm cm^{-3}}$.
The spin Hall effect was detected via the optical technique of
Kerr rotation microscopy
in the presence of an external magnetic field in a Hanle-type setup.
The Kerr rotation signal as a function of the applied magnetic field
could be fitted well by a Lorentzian where the spin lifetime $\tau_{s}$ entered as
a fit parameter. By scanning over the sample, the authors
obtained spatial profiles of the spin density
along the direction perpendicular to the applied electric field. Fitting
this data to solutions of the spin diffusion equation, the spin diffusion
length was extracted. The values for this quantity lie between two and four
micrometers and are, within the error bars, insensitive to the
electric field varying between zero and $25{\rm mV}\mu{\rm m^{-1}}$. Combining
this data with results for the spin lifetime, the authors inferred
a spin conductivity of about $0.5\Omega^{-1}{\rm m^{-1}}$, where the latter
quantity  was converted to units of
charge transport by multiplying with a factor of $e/ \hbar$.

The samples investigated by Kato {\em et al.} are n-doped and
clearly in the bulk regime.
Therefore, the intrinsic spin-orbit coupling to conduction-band electrons
is dominated by the bulk Dresselhaus term (\ref{bulkdressel}) which was
studied by Bernevig and Zhang as a possible intrinsic
mechanism for spin Hall transport.
\cite{Bernevig05d}. The authors
apply their results to the experiments by Kato {\em et al.} \cite{Kato04}.
However, the agreement between theoretical predictions and the
experimental findings is certainly not very convincing. Moreover,
Kato {\em et al.} find only a very negligible dependence of the
spin Hall transport on strain applied to the system. This observation also
strongly disfavors an intrinsic mechanism. In fact, Engel, Halperin, and
Rashba \cite{Engel05}
have developed a theory of extrinsic spin Hall transport in GaAs
based on impurity scattering and found reasonable agreement with
the results by Kato {\em et al.} \cite{Kato04}.

Further results on this type of experiments were reported by 
Sih {\em et al.} from the same research 
group for the case of n-doped GaAs quantum wells \cite{Sih05} (as opposed
to the bulk epilayers discussed so far). The quantum well here has a width of
$75\AA$ with a sheet density of $n=1.9\cdot10^{12}{\rm cm^{-2}}$ and a mobility
of $\mu=940{\rm cm^{2}/Vs}$. From the analysis of their Kerr rotation data
Sih {\em et al.} conclude that the signatures of spin Hall transport
seen in their experiment are also most likely due to an extrinsic mechanism. 

We now turn to spin Hall transport of holes.
Wunderlich{\em et al.} have investigated spin Hall effect in a p-doped
triangular quantum well which is part of a p-n junction light emitting diode
\cite{Wunderlich05}. The quantum well has a sheet hole density of 
$n=2\cdot10^{12}{\rm cm^{-2}}$ which is for the still low enough, for the given
sample geometry, such that only the first heavy hole subband is occupied.
Thus, the intrinsic spin-orbit coupling to these heavy holes 
can be expected to be governed by the Hamiltonian (\ref{defham}),
leading to a spin Hall conductivity given by Eq.~(\ref{fullspinHall}).
The spin accumulation at the edges of the well is detected by the
circular polarization of the light emitted from the diode. In a subsequent
publication, the authors presented further details on this technique
\cite{Kaestner05}. In summary, Wunderlich {\em at al.} conclude
that the spin Hall transport seen in their results is most likely of
intrinsic nature, i.e. due to Rashba spin-orbit coupling as
described by the Eq.~(\ref{defham}), although further work is
needed to fully establish this conclusion \cite{Wunderlich05}. 

Let us now mention some proposals for further possible experimental
investigations of spin Hall transport. Hankiewicz {\em et al.} have considered
an H-shape two-dimensional electron system using the Landauer-B\"uttiker 
formalism \cite{Hankiewicz04}. The sample consists of two, say, horizontal
bars connected in their centers by a shorter vertical bar. A voltage applied
along one of the long bars should drive a spin current through the 
shorter connecting bar into the other horizontal bar. There this spin current
should manifest itself by a voltage along the bar which can be 
calculated by inverting the spin conductivity tensor. Numerical calculations
support the feasibility of this experimental approach \cite{Hankiewicz04}.

A scheme to determine the spin Hall conductance purely via measurements
of charge transport was put forward by Erlingsson and Loss 
\cite{Erlingsson05b}.
The authors study a planar four-terminal setup.
Using conventional scattering formalism, they express the spin Hall
conductance in terms of voltages, charge conductances, and charge current noise
quantities. This allows, in principle, to infer the spin Hall conductance
just from electric measurements, avoiding the necessity of any magnetic or
optical element \cite{Erlingsson05b}. 

\section{Conclusions and outlook}
\label{concl}

We have given an overview on recent theoretical and experimental 
developments concerning spin Hall transport in semiconductors. This phenomenon
has certainly been over the last years
one of the most intensively worked on topics in the solid-state community,
and research efforts still continue to grow.

The theoretical situation
regarding spin Hall effect in the two-dimensional electron gas with
spin-orbit coupling linear in the momentum is by now well settled: There
is no spin Hall effect in an infinite system if any kind of dissipation
mechanism is present. This is, however, not a statement about finite 
mesoscopic systems, and further both theoretical and experimental work
is to be expected here. On the other hand, p-doped quantum wells, i.e.
the two-dimensional hole gas, appears to be a particularly attractive 
system. 

A challenge for future theoretical work is certainly to
extend many present results to the case of finite temperature, and, more
importantly, include the Coulomb interaction between charge carriers.
The last point is addressed by only very few papers so far
\cite{Shekhter05,Dimitrova04a}. A further recent theoretical 
development are studies on the {\em zitterbewegung} of electron and hole
wave packets due to spin-orbit interaction in semiconductors
\cite{Schliemann05b,Schliemann05c}. These systems offer the possibility
to experimentally detect the relativistic phenomenon of {\em zitterbewegung} 
which appears to be quite inaccessible in the case of free electrons
\cite{Schliemann05b,Schliemann05c}.

Future challenges for experiments on spin Hall transport include
the clarification and discrimination of extrinsic and intrinsic mechanisms.
Moreover, from both a theoretical and an experimental point of view, it
is undoubtedly desirable to develop setups and techniques which allow to
detect spin Hall transport independently from spin accumulation. 

\section*{Acknowledgements}
We thank E.~S. Bernardes, C. Bruder, D. Bulaev, G. Burkard, O. Chalaev, 
M. Duckheim, J.~C. Egues, S.~I. Erlingsson, M. Lee, D. Loss, D. Saraga, 
and R.~M. Westervelt for fruitful collaboration 
and/or discussions on effects of spin-orbit coupling in semiconductors.
This work was
supported by the SFB 689 ``Spin Phenomena in reduced Dimensions''.

\end{document}